# Investigation of Cross-border Banking Activities

**Hoang Anh Le (b) Xuan Vinh Vo (a),**

[a] School of Banking, University of Economics, Ho Chi Minh City and CFVG Ho Chi Minh City, Vietnam.

[b] College of Foreign Economic Relations, Ho Chi Minh City, Vietnam.

## Abstract

This paper investigates cross-border lending behavior from the Group of Seven (G7) during the 2001 - 2013 period. We employ gravity model to consider how bilateral factors, global factors, and other determinants of pull factors affect cross-border lending. The empirical results demonstrate that driving factors for cross-border lending have been changing since the 2008 Global Financial Crisis. Particularly, continent variable has more significant correlation with cross-border claims during the post-crisis period, while distance variable becomes less important during that time than it was in the pre-crisis period. Additionally, higher lending claims is more likely related to common language after the financial crisis. Moreover, the role of pull factors, except the size of borrowing economies, is not significant in explaining cross-border banking activities since GFC.

*Keywords: Cross-border banking, international banking, gravity model, G7*

## 1. Introduction

The recent financial crisis severely affects the banking activities. Firstly, many papers report that cross- border lending decreases around the time of crisis (Bremus & Fratzscher, 2015; Popov & Udell, 2012). Secondly, this reduction is likely driven by economic geography. Herrmann and Mihaljek (2013), for example, assert that the Central and Eastern Europe is less affected by the decrease of foreign claims compare to other emerging countries are due to its stronger financial and monetary ties with creditors. These facts raise a question if there are any change in international banking behaviors after crisis.

Many authors express concerns about the home bias effects which may play an important role along with the Global Financial Crisis (GFC). Especially, home bias effects are evident in cross-border banking in the Euro area (Bremus & Fratzscher, 2015; Heuchemer, Kleimeier, & Sander, 2009). This implies that international banks focus more on their familiar and closer market. Given these trend, what drive cross-border lending from a country to another continue to be an important topic that attracts interest from many scholars (Müller & Uhde, 2013; Siregar & Choy, 2010; Temesvary, 2014).

In this paper, we attempts to shed further light on what factors influence the international lending behavior of G7 countries by using the gravity equation approach. There are two separate models employed in our papers. The former focuses on testing the economic relations, financial development, culture and political similarities and differences employing bilateral data. The latter examines pull factors including size of recipient countries, governance risk and domestic banking activities.

Our main finding will contribute to studies in the field of international capital liberalization and bank risk management. Obviously, the data of cross-border lending shows that there is a strong correlation between the financial crisis and G7 lending behavior. More than a half of G7 is European countries, which provide more credits to countries located in the same continent.

This paper supports the regulation in international financial market. Our data clearly show how much cross-border lending is concentrated in specific economic area. Understanding how financial development, government quality and public debt drive the lending from G7 to the borrowing countries, government will adjust the financial policy of borrowing countries to enrich the financial development.

The remainder of the paper proceeds as follows. Section 2 reviews the literature on cross-border banking activities. Section 3 presents the models, describes the data, and introduces variables and estimation methods. Section 4 presents the empirical results. Section 5 concludes the paper.

## 2. Literature Review

Cross-border lending is a major of international capital flows and plays an important role in international economics. There are many existing papers investigating cross-border lending behavior. Some studies document two main motivations of lenders which are averting of risk and preferring lower transaction costs, which are usually implied by the characteristics of borrowing countries and lending countries (Herrmann & Mihaljek, 2013; Siregar & Choy, 2010; Temesvary, 2014). Other studies give evidence for the home bias effects which is significant driver of international financial investment (Bremus & Fratzscher, 2015; De Haas & Van Horen, 2013; Heuchemer et al., 2009).

Regarding risk composited by the borrowing country, the quality of governance is a significant factor in explaining international banking flows from developed world to East Asian (Siregar & Choy, 2010). This study states that political instability, weakness in legal, judicial, and bureaucratic system are all strong determinants of cross-border banking after the GFC.

Credit risk may also related to asymmetric information, which is considered to have a strong impact on cross-border lending behavior. Many variable can be used as proxies for information disadvantage. First, the distance between the lending country and the borrowing country is usually considered as an indicator for asymmetric information. Some papers show that there is a negative correlation between distance and international financial flows (Herrmann & Mihaljek, 2013; Martin & Rey, 2004; Siregar & Choy, 2010). We may assume that advance in technology is not always consistent with available information. Therefore, the distance remains its role in the international capital flows. However, others studies find that the role of distance has been less important (Buch, 2005; De Haas & Van Horen, 2013). Second, the trading history between lending country and borrowing country also indicates for the asymmetric information. As a consequence, trading partners which are more interactive with lending country usually attracts higher credit. The existing literature finds many evidences that physical trading in goods and services embolden financial trading (Heuchemer et al., 2009; Martin & Rey, 2004; Siregar & Choy, 2010).

A number of papers attribute the home bias effects to explain cross-border banking. Heuchemer et al. (2009), for example, finds that those countries which are more similar with creditors in financial development tend to own more cross border claims than others do. In additional, cultural differences may discourage the amount of cross-border lending (see also De Haas & Van Horen, 2013). Most recently, Bremus and Fratzscher (2015) conclude that home bias increases rapidly in the cross-border claims, especially among banks in the Euro area.

3. **Research methodology and data**

   *3.1. Models*

This section presents a brief description of the models. Following Heuchemer et al. (2009), our first model is the gravity one which is presented as follows:

$$\ln Claim_{ijt} = \alpha_0 + \beta_1 * \ln Size_{ijt} + \beta_2 * \ln Rel_{ijt} + \beta_3 * \ln Similar_{ijt} + \beta_4 * \ln Dist_{ijt} + \beta_5 * Border_{ij} + \sum \beta_k * X_{ijt} + \mu_{ijt}$$

where

$Claim_{ijt}$ is the bilateral claim, which presents for the amount of foreign claim outstanding from lending country i to borrowing country j at the end of the year t;

*Size* measures the economies scale of both lender economy and borrower economy. This variable is calculated similar to Heuchemer et al. (2009):

$$Size_{ijt} = \ln(GDP_{it} + GDP_{jt})$$

where $GDP_{it}$ and $GDP_{jt}$ are the gross domestic product (in billions of USD) of lending country i and borrowing country j in year t, respectively. The data for GDP are taken from World Bank's World Development Indicators WDI. This indicator, *Size*, is expected to have a positive effect on bilateral claims.

$Rel_{ijt}$ and $Similar_{ijt}$ respectively measure the differences and similarities in economic sizes between those creditors and debtors. These are also calculated following Heuchemer et al. (2009) formula:

$$Rel_{ijt} = \left| \ln\left(\frac{GDP_{it}}{Population_{it}}\right) - \ln\left(\frac{GDP_{jt}}{Population_{jt}}\right) \right|$$

$$Similar_{ijt} = \ln(1 - \left(\frac{GDP_{it}}{GDP_{it}+GDP_{jt}}\right)^2 - \left(\frac{GDP_{jt}}{GDP_{it}+GDP_{jt}}\right)^2)$$

where $GDP_{it}$ and $GDP_{jt}$ are the gross domestic product (in billions of USD) of lending country i and borrowing country j in time t, respectively; $Population_{it}$ and $Population_{jt}$ are the population of country i and country j in time t, respectively. Data for these variables are collected from the IMF. A negative value coefficient of Rel would indicate that the more similar in the GDP per capita leads to higher claims trading. Higher value of Similar, by contrast, implies more similarities between lenders and borrowers. This variable is expected to have positive sign in the model.

For further understanding the cross-border lending behavior, we also consider the impact of financial development on bilateral cross-border claims (Heuchemer et al. 2009) . This study suggests that the national banking market conditions may be likely to impact more on bilateral loans than the economic conditions. The relative financial development and overall similarities of the size of financial sector are measured by four indicators including credit, deposit, money and market. First, credit is defined as domestic credit to private sector by banks, as following IMF's definition. Second, deposit is described as demand, time and saving deposits in deposit money banks as a share of GDP, calculated using the following deflation method. Third, money as money supply is often called M2, which is also following IMF' definition and calculated method. Fourth, market is the value of stocks traded to measure the stock market. Data are collected from the World Bank's WDI database. Furthermore, we construct four bilateral variables as follows:

$$REL_{FDijt} = \left|\ln(FD_{it}) - \ln(FD_{jt})\right|$$

$$SIMILAR_{FDijt} = Ln\left(\frac{1 - (FD_{it} * GDP_{it})}{(FD_{it} * GDP_{it} + FD_{jt} * GDP_{jt})^2}\right)$$

$$+ Ln\left(\frac{1 - (FD_{jt} * GDP_{jt})}{(FD_{it} * GDP_{it} + FD_{jt} * GDP_{jt})^2}\right)$$

where FD is financial development indicators.

*Dist$_{ij}$* is the distance between the capital city of lending country i and the capital city of borrowing country j. According to the traditional gravity theory, the potential flows of sources including labor, goods or other factors are decreased by the distance between suppliers and demanders (Anderson, 2011). During crisis, international banks tend to reduce credit to borrowing countries. Credit reduction is relatively associated with geographical features. They assert that longer distance may cause higher cost of lending, especially, during crisis when banks must monitor their loans more carefully. For example, Ruckes (2004) states that it is more expensive for creditors to manage their claims offering to remote borrowers. Furthermore, De Haas and Van Horen (2013) also find that advanced economies tend to provide more loan to geographically closed emerging economies which financial relationships and monetary links are tied to host countries. The distance between borrower and lender is collected from the timeanddate.com website. This factor is expected to have a negative impact on foreign claims.

*Border$_{ij}$* is a dummy variable, taking the value of 1 if country i and j have common border and 0 otherwise. This variable is a measure of transaction cost. Countries which have common border would be expected to reduce transaction cost. Therefore, this dummy variable is expected to have a positive sign.

$X_{ij}$ is a vector contain set of factors which measures cultural and political similarities between lending countries and borrowing countries. These variable are language, culture and political indicators.

*Language*: we define two countries in common language if they are in common official or main language or widely understood language, which is a proxy for cultural factor. Dataset is collected from the world fact book from Central Intelligence Agency (CIA). Although English is currently popular in international trading and international lending, difference in language plays as a barrier of cross-border claims because it still raises cost for creditors to understand the business announcements in other language. In fact, there are not all information available in English in every country, and, in some cases, those in local language are essential information for making lending decision.

*Culture*: We use the data from six cultural dimensions of Hofstede[1], which are power distance, individualism, masculinity, uncertainly avoidance, long term orientation and indulgence.

*Political indicators*: Following Heuchemer et al. (2009), this paper investigates political drivers by using six time- varying dimensions of governance defined by the World Bank: voice and accountability (VOICE), political stability and absence of violence (POLSTAB), government effectiveness (GOVEFF), regulatory quality (REGQAL), rule of law (LAW), and control of corruption (CORRUP). We then calculate an overall political risk proxy as measure of political dissimilarity, following the Euclidean distance formulation below.

Culture and Political indicators enter the regression as calculating by Euclidean distance:

---

[1]Source: https://geert-hofstede.com/countries.html?culture1=86&culture2=18Appeal

$$ED_{ij} = \sqrt{\sum_{k=1}^{k}(V_{ik} - V_{jk})^2}$$

where ED is the Euclidean distance and V are the different proxies. The higher ED implies the more difference between the countries involved, which is expected to have a negative sign.

Two additional explanation variables including bilateral trade between the G7 and the destination countries (*Lntrade$_{ijt}$*) and Bilateral Foreign Direct Investment (*FDI$_{ijt}$*) are employed in our first model. These variables are used following Heuchemer et al. (2009). The bilateral trade in our regression is computed as the total of import and export between lender and borrower countries in year t. These variables enter the regression in natural logarithm form. Higher trade volume implies a closer relationship between home and host economies. Lenders would likely lend money to those who is more familiar. In fact, the more interaction with each other leads to the lower restriction in the information cost. Predictably, trade in goods and service encourages more trade in financial asset between two countries. Following with a strong relationship in physical trading, lender can use the accumulated information in trading goods for making lending decisions. Therefore, this variable is expected to have a positive impact on cross-border lending. Bilateral import and export are collected from the Trade Map[2].

Similarly, the higher amount of FDI flows from lending countries to borrowing countries the more information about the debtors is recorded. In additional, lender countries prefer to finance the business that is closely linked to their home economy, especially in difficult time. As a consequence, this variable is expected to have a positive sign. The data for FDI are collected from UNCTAD FDI/TNC database.

---

[2] Trade Map provides indicators on export performance, international demand, alternative markets and competitive markets, as well as a directory of importing and exporting companies. Trade Map covers 220 countries and territories http://www.trademap.org/

For further understanding of the determinants of cross-border claims, we add three more variables into our first model. Those variables are used to indicate the home bias effects, the exchange rate risks and global risks.

$$\ln Claim_{ijt} = \alpha_0 + \beta_1 * \ln Size_{ijt} + \beta_2 * \ln Rel_{ijt} + \beta_3 * \ln Similar_{ijt} + \beta_4 * \ln Dist_{ij} + \beta_5 * Border_{ij} + \beta_6 * Continent_{ij} + \beta_7 * ER_{ijt} + \beta_8 * VIX_t + \sum \beta_k * X_{ijt} + \mu_{ijt}$$

where

*Continent$_{ij}$* is a dummy variable, taking the value of 1 if home country and destination country are in the same continent and 0 otherwise. Similar to *Dist,* this variable is a proxy for transaction cost. Moreover, those countries in the same continent usually have strong interaction with each other and they will also have the similar history and development which lead to be more similar culture. Therefore, this dummy variable is also considered as a proxy for culture similarities;

*ER$_{ijt}$* proxy for volatility of bilateral exchange rate between currency of country i and currency of country j in year t. The data are collected from OANDA. This variable is a measure of exchange rate risk, which is calculated as the standard deviation of monthly logarithmic change of exchange rate, following the formula seen in Nedzvedskas and Aniūnas (2007).

$$ER_i(t) = \begin{cases} \ln(\frac{X_i(t)}{X_i(t-1)}) \\ \ln(\frac{1}{X_i(t-1)}) & if\ X_i(t) = 0 \\ \ln X_i & if\ X_i(t-1) = 0 \end{cases}$$

The higher figure of ER indicate that the exchange rate exposure involved in the cross-border claims is very high. This variable is expected to have a negative sign in the model;

$VIX_t$ is the natural logarithm of CBOE Volatility index in year t. We employ the CBOE index to measure the global risks. Rational investors or lenders tend to be less motivated if risk increases. This variable is expected to have negative correlation with bilateral claim.

In the second model, we will investigate the role of borrowing country characteristics in pulling cross-border claims from G7, which is presented as below:

$$lnT\_Claim_{jt} = \alpha_0 + \beta_1 * lnGDPjt + \beta_2 * \ln Export_{jt} + \beta_3 * \ln Import_{jt} + \beta_4 * DEBT_{jt} + \beta_5 * Gov_{jt} + \beta_6 * Credit_{jt} + \varepsilon_{it}$$

where T_Claim$_{jt}$ is total claims from G7 to country j at year t; GDP$_{jt}$ is the gross domestic product of borrowing country j in year t; Export and Import are the total export and import, respectively, between country j and all G7 countries in year t; DEBT$_{jt}$ is the general government gross debt per GDP ratio of host countries in year t; Gov$_{jt}$ is the governance indicator of the borrowing country j in year t; Credit$_{jt}$ is the domestic credit to private sector by banks (per cent of GDP) of country j in year t.

- *Gross domestic product (GDP)*: we employ the sum of gross domestic product value added as a proxy of economic scales of borrowing countries. The data are collected from the World Bank's WDI. This variable enters our regression in natural logarithm form. It is expected to have a positive impact on foreign claims inflowing an economy.

- *Trade (Export – Import):* These variable are proxies for the openness degree of the borrowing economies. The more interaction with G7 is, the higher amount of claims flows in. Raw data are collected from the Trade Map. These variables enter the regression in natural logarithm form and are expected to appear with positive signs.

- *General government gross debt (DEBT):* This variable is another factor which reflects the domestic condition of recipient country. Government which responses to higher level of debt will contribute to higher country-specific risk. Laubach (2010) finds the positive coefficient between the country yield spread and the level of debt to GDP ratio. This result implies that higher government debt likely leads to higher risk premium or default risks. In our model, we employ general government gross debt by using the percentage figures given on the World Bank's WDI data. This variable is also expected to have negative impact on cross-border lending.

- *Governance (GOV)***:** this variable is a crucial factor for cross border lending. Thus, we use governance indicator similar to Kaufmann, Kraay, and Mastruzzi (2004), as mentioned above. The higher scores rated would be likely to expect that higher claims inflowing.

- *Credit***:** This variable is the percentage of domestic credit to private sector by banks of GDP. Data are collected from World Bank's WDI. This variable is a proxy of financial system development in borrowing countries: a lower rate implies a lower banking activity. Reasonably, bank system development encourages cross-border claims inflow. Therefore, it is expected to have positive sign.

*3.2. Data description*

Foreign claims from G7 to different countries are taken from Bank International Settlements (BIS) Consolidated banking statistics. Figure 1 to figure 8 give an overview of bank flows during the period between 2001 and 2013.

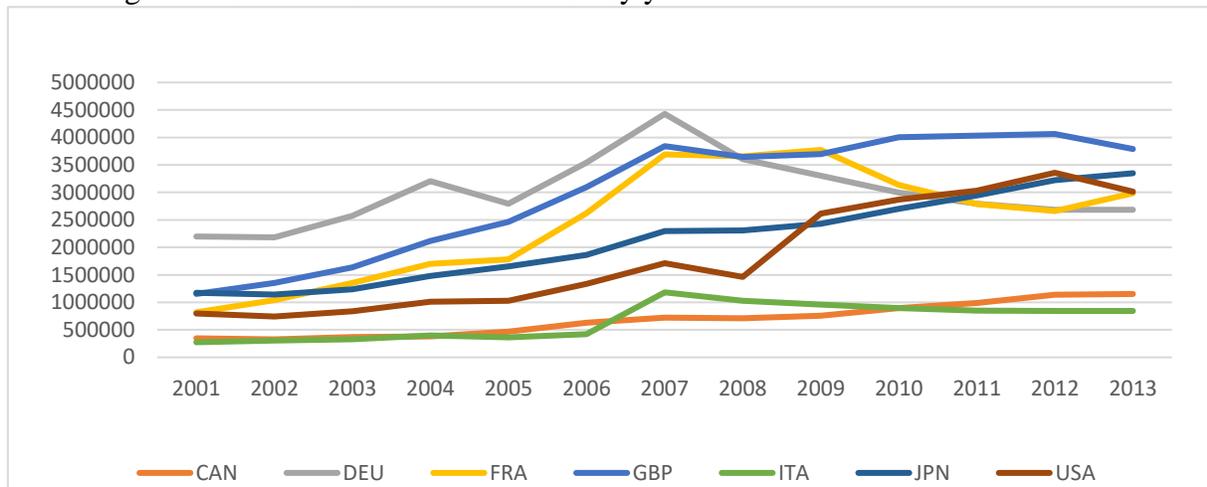

Figure 1. Cross-border claims from G7 by year

Source. BIS – Consolidated Banking Statistics

Overall, it is apparent that foreign claims from G7 to the rest of the world increase considerably between 2001 and 2013. However, European lenders are likely to decrease their lending significantly after financial crisis while other countries in G7 continue to be expanded theirs. Especially, German holds the highest amount before 2007 and then their claims to all countries as well as developed countries drop dramatically as followed. Similarly, claims of France and Italy share a similar pattern with German, which increase rapidly between 2006 and 2007 and go down during the next period. In contrast, there is different trend in claims of Japan, an Asian country, which witness a gradual increase before 2007, then decrease very slightly during 2008, and immediately the figure turned to its upward trend for the following years. Another non-European country is Canada, which holds the smallest amounts of foreign claims compare to others in G7, experience a gradual increase for the whole period without any drop during the crisis.

The fluctuation between these countries' cross-border lending could be explained by the different destinations. Figure 2 to figure 8 show the bilateral claims from each of the G7 to different continents. Most of the G7 seem to concentrate their claims on the economic area which has closed link to lender countries. For instance, European Banks (French, German and especially Italian

Bank) offer the largest proportion of total claims in Europe and secondary in North America, while Canadian Banks provided more claims to North America and Europe area, which consist about one third of total claims before 2006 and its share is curtailed in the following years. By contract, the UK, the USA and Japan extend more claims to countries which are not in the same continent. In fact, the UK lend more to North America, Europe and Asia. Regarding the USA, the European countries are the largest debtors and the other major borrowing countries belong to Asia and North America. Lastly, Japanese banks lend more to North America and Europe.

Figure 2. Claims from Canada to different continents

Source. BIS – Consolidated Banking Statistics; Own calculations

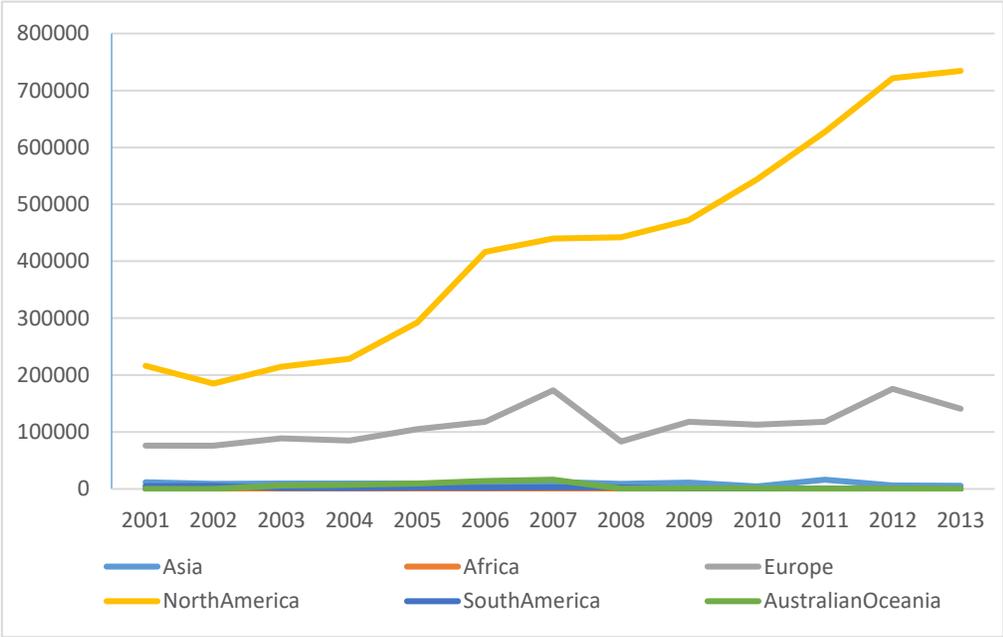

Figure 3. Claims from German to different continents

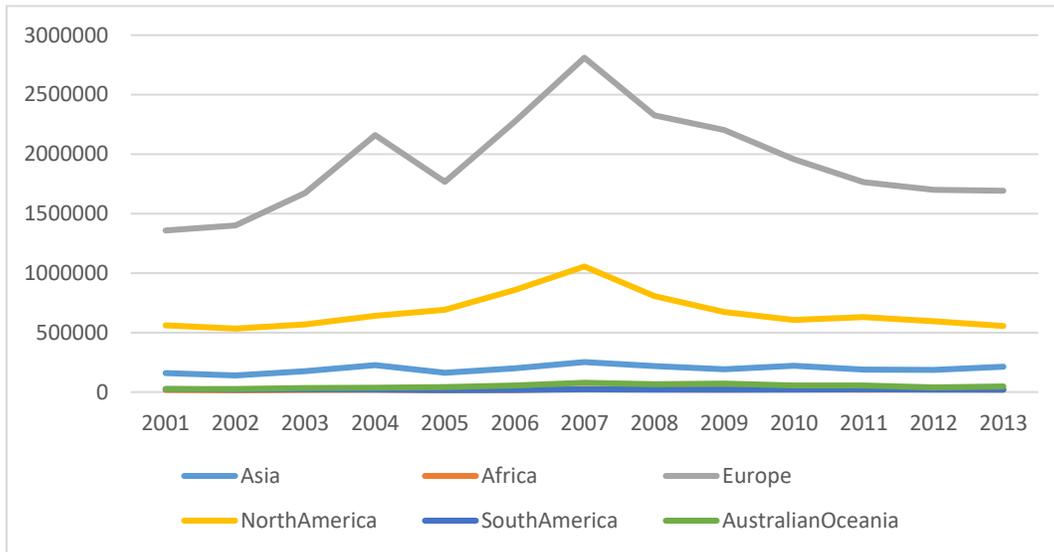

Source. BIS – Consolidated Banking Statistics; Own calculations

Figure 4. Claims from France to different continents

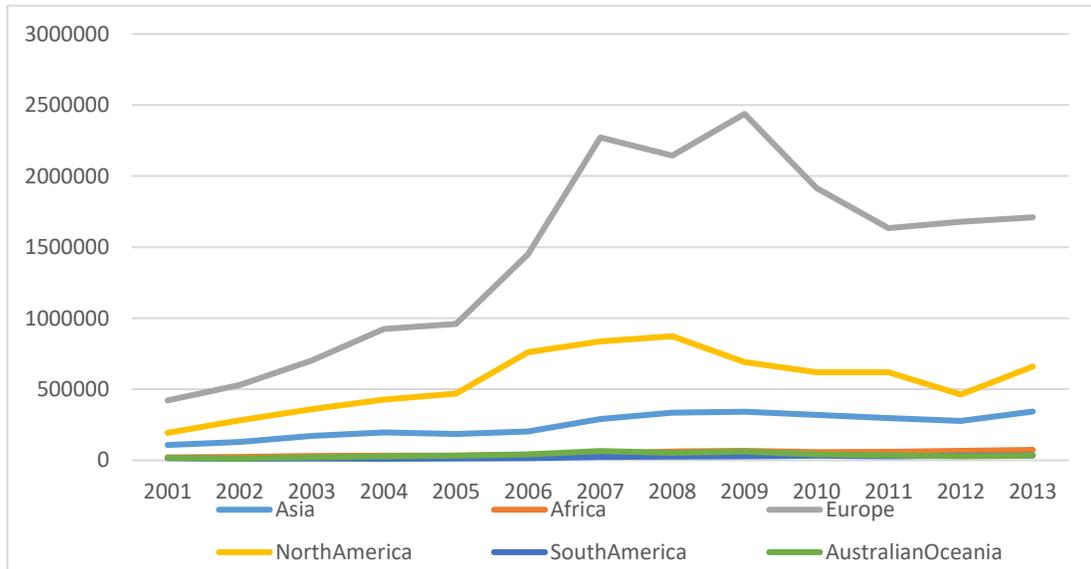

Source. BIS – Consolidated Banking Statistics; Own calculations

Figure 5. Claims from UK to different continents

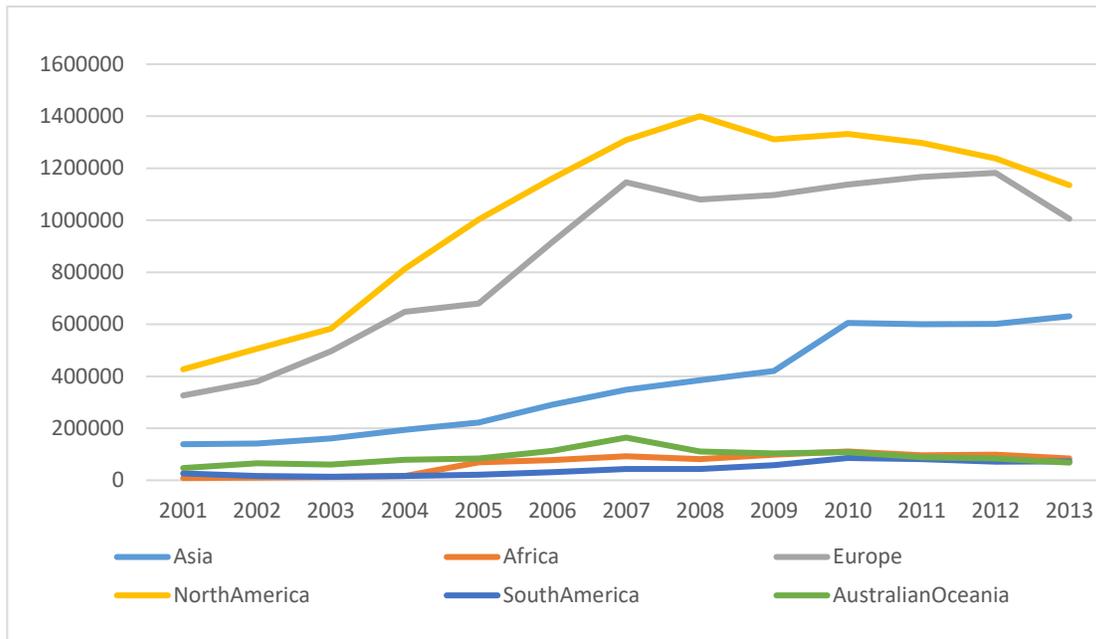

Source. BIS – Consolidated Banking Statistics; Own calculations

Figure 6. Claims from Italia to different continents

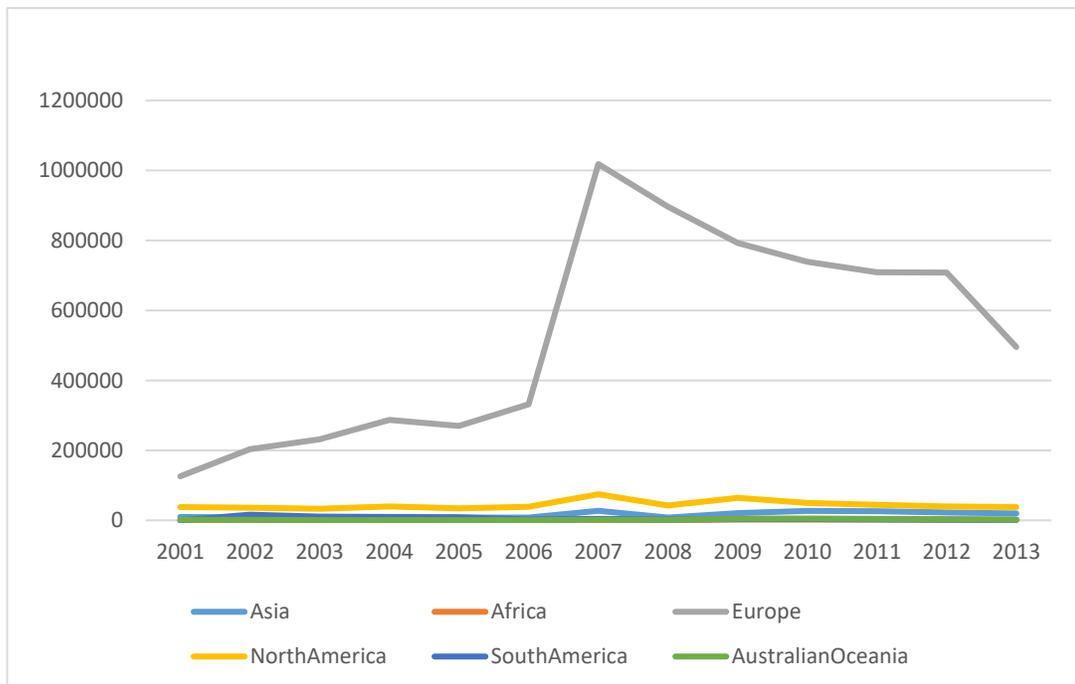

Source. BIS – Consolidated Banking Statistics; Own calculations

Figure 7. Claims from Japan to different continents

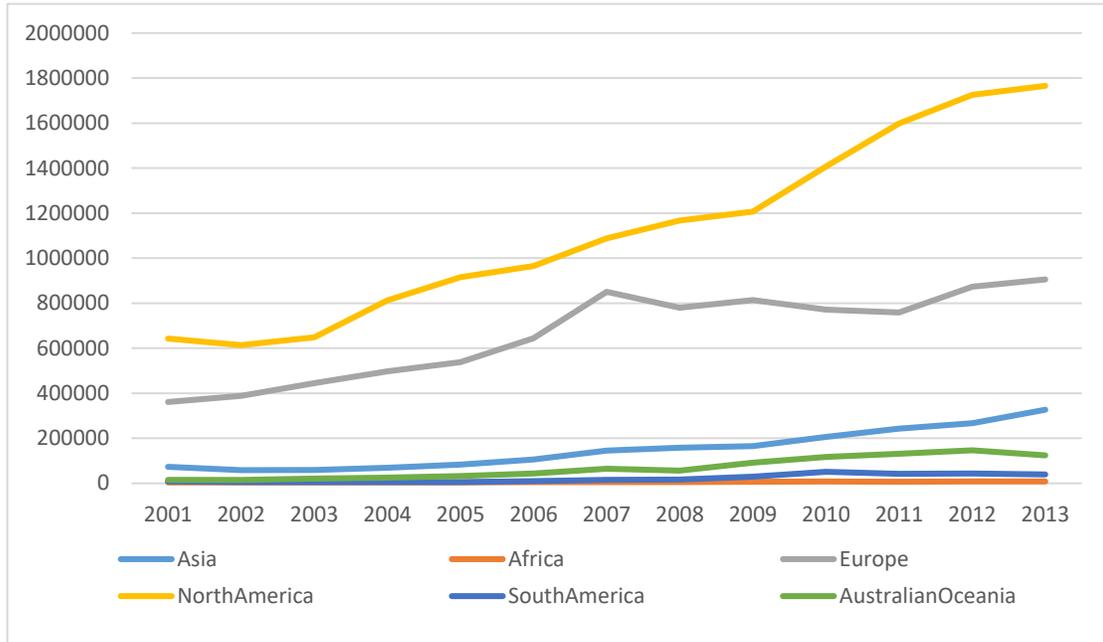

Source. BIS – Consolidated Banking Statistics; Own calculations

Figure 8. Claims from USA to different continents

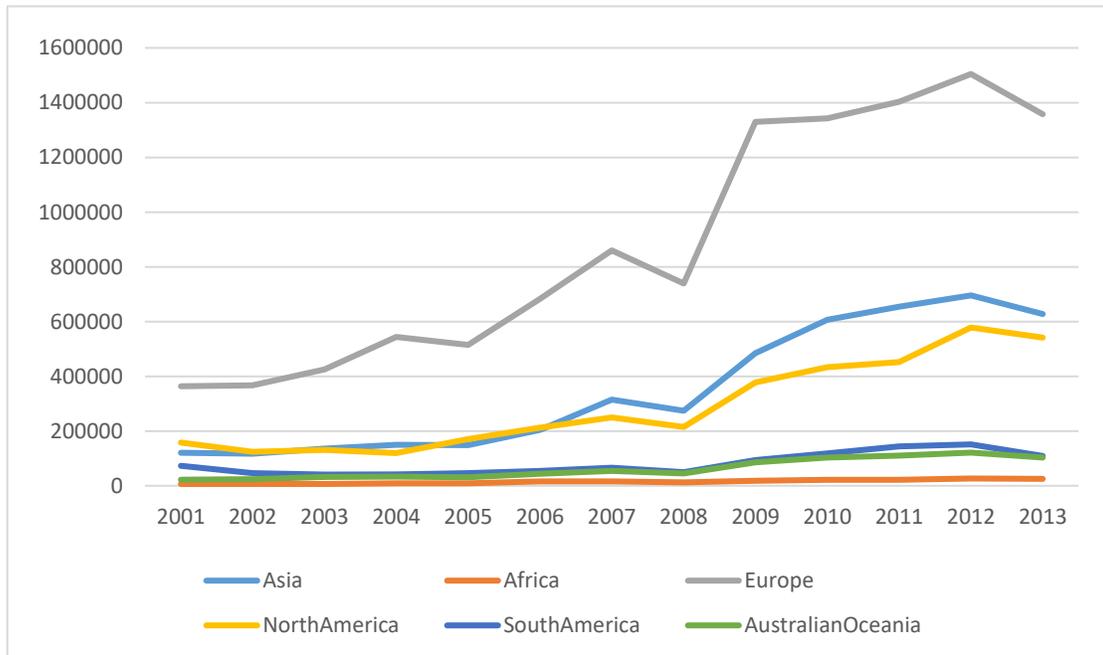

Source. BIS – Consolidated Banking Statistics; Own calculations

**Table 1 Descriptive statistics**

|  | Observations | Mean | Std. Dev. | Min | Max |
| --- | --- | --- | --- | --- | --- |
| Lnclaim | 9385 | 6.2721 | 3.1152 | 0.0000 | 14.0177 |
| Lnsize | 17430 | 3.3597 | 0.0247 | 3.3205 | 3.4338 |
| Rel | 15661 | 10.214 | 0.6366 | 1.9742 | 10.7256 |
| Similar | 17430 | -4.575 | 2.2982 | -12.6159 | -0.6932 |
| Rel_credit | 15228 | 1.4566 | 0.9884 | 0.0000 | 5.9413 |
| Similar _credit | 14993 | -5.841 | 2.8523 | -14.7636 | -0.6932 |
| Rel _m2 | 15062 | 1.0787 | 0.6697 | 0.0000 | 4.2535 |
| Similar _m2 | 14937 | -5.438 | 2.5673 | -12.1014 | -0.6932 |
| Rel _deposit | 12252 | 1.1076 | 0.7780 | 0.0000 | 4.7224 |
| Similar _deposit | 12064 | -5.802 | 2.2758 | -13.0657 | -0.6932 |
| Rel _market | 9156 | 3.1941 | 2.3969 | 0.0000 | 12.0805 |
| Similar _market | 8720 | -15.7476 | 2.6938 | -25.7182 | -9.2276 |
| Lndist | 19006 | 8.6866 | 0.7867 | 4.4067 | 9.8512 |
| Border | 19110 | 0.019 | 0.1367 | 0.0000 | 1.0000 |
| Continent | 19110 | 0.2156 | 0.4113 | 0.0000 | 1.0000 |
| Pol | 16226 | 3.6706 | 1.9241 | 0.0000 | 10.0803 |
| Lntrade | 16377 | 12.619 | 2.9318 | 0.0000 | 20.2721 |
| LnFDI | 5093 | 6.1228 | 2.9517 | -0.7103 | 12.9115 |
| Language | 19019 | 0.3240 | 0.4680 | 0.0000 | 1.0000 |
| Culture | 9100 | 64.329 | 20.8276 | 0.0000 | 124.2014 |
| Exchange | 14116 | 0.8764 | 0.7338 | 0.0000 | 7.5616 |
| *LnVIX* | 19110 | 3.0094 | 0.3131 | 2.5500 | 3.4871 |
| *LnGDP* | 2545 | 23.5541 | 2.3999 | 16.8932 | 31.6428 |
| *LnExport* | 2366 | 14.0814 | 3.0747 | 4.9558 | 22.6411 |
| *LnImport* | 2366 | 14.3356 | 2.6024 | 7.2779 | 22.4910 |
| *Debt* | 2391 | 53.8228 | 53.4901 | 0.0000 | 789.8330 |
| *GOV* | 2327 | -0.0178 | 0.9942 | -2.4500 | 2.4297 |
| *Credit* | 2313 | 46.4184 | 42.1835 | 0.1859 | 311.7775 |

## 4. Empirical Results

The table 4.1 shows that G7 countries lend more money to those countries which are more different in financial development. However, they lend more in countries with similar in economic size, same continent and political conditions. This is consistently with the expectation based on home bias effects. In term of risk aversion, bilateral exchange rate volatility has negative effect on claims from G7 to their destinations, with significant statistically (at the 1% level). However, some factors appear in our result with unexpected signs. First, distance variable has a positive correlation with significant statistically although continent variable has a positive correlation. Second, global risk has a positive correlation with cross-border claims form G7, which is also different with the expectation base on risk aversion. Another interesting point, cultural factor is less significant correlation with cross-border lending. Variable which present cultural differences, especially, appears with statistical significance in almost the regression results. Although, language variable is statistically significant when we use market, m2 and credit proxy for similar and relative.

**Table 4.1 Regression results for the full sample**

|  | PROXY FOR SIMILAR AND REL | | | | |
| --- | --- | --- | --- | --- | --- |
| Variable | GDP | CREDIT | M2 | MARKET | DEPOSIT |
| _cons | -104.3182*** | -90.8821*** | -92.8481*** | -55.7532*** | -62.2375*** |
|  | (0.0000) | (0.0000) | (0.0000) | (0.0000) | (0.0000) |
| size | 30.7240*** | 26.4810*** | 27.0703*** | 15.4249*** | 16.8922*** |
|  | (0.0000) | (0.0000) | (0.0000) | (0.0000) | (0.0000) |
| Rel | -0.1017** | | | | |
|  | (0.0220) | | | | |
| similar | 0.5980*** | | | | |
|  | (0.0000) | | | | |
| rel _credit | | 0.2570*** | | | |
|  | | (0.0000) | | | |
| similar_credit | | 0.3649*** | | | |
|  | | (0.0000) | | | |
| rel _m2 | | | 0.4232*** | | |
|  | | | (0.0000) | | |
| similar _m2 | | | 0.4691*** | | |
|  | | | (0.0000) | | |
| rel_market | | | | 0.0630** | |
|  | | | | (0.0620) | |
| similar _market | | | | 0.0865** | |
|  | | | | (0.0050) | |
| rel _deposit | | | | | -0.2626** |
|  | | | | | (0.0030) |
| similar _deposit | | | | | -0.1401** |
|  | | | | | (0.0030) |
| lndist | 0.3573** | 0.1965* | 0.2136** | 0.1744* | 0.3097** |
|  | (0.0040) | (0.0590) | (0.0380) | (0.0810) | (0.0080) |
| Border | -0.5069 | -0.0591 | -0.1561 | -0.0775 | 0.1177 |
|  | (0.2270) | (0.8380) | (0.5880) | (0.7790) | (0.7050) |
| Continent | 1.0831*** | 0.7132** | 0.8415*** | 0.3257 | 0.9826*** |
|  | (0.0000) | (0.0010) | (0.0000) | (0.1160) | (0.0000) |
| pol | -0.1759*** | -0.2110*** | -0.2070*** | -0.2089*** | -0.2203*** |
|  | (0.0000) | (0.0000) | (0.0000) | (0.0000) | (0.0000) |
| lntrade | 0.4736*** | 0.5475*** | 0.5299*** | 0.7268*** | 0.6115*** |
|  | (0.0000) | (0.0000) | (0.0000) | (0.0000) | (0.0000) |
| lnFDI | 0.1304*** | 0.1402*** | 0.1389*** | 0.1426*** | 0.1250*** |
|  | (0.0000) | (0.0000) | (0.0000) | (0.0000) | (0.0000) |
| language | 0.2246 | 0.4901** | 0.4779** | 0.5240*** | 0.2788 |
|  | (0.2260) | (0.0010) | (0.0020) | (0.0000) | (0.1100) |
| culture | 0.0026 | -0.0013 | 0.0012 | -0.0031 | -0.0071* |
|                      | (0.4970) | (0.6960) | (0.7060) | (0.3180) | (0.0560) |
| ER                   | -0.4044*** | -0.3653*** | -0.3715*** | -0.4320*** | -0.2938*** |
|                      | (0.0000) | (0.0000) | (0.0000) | (0.0000) | (0.0000) |
| lnVIX                | 0.1390*** | 0.1055*** | 0.1351*** | 0.1214*** | 0.1619*** |
|                      | (0.0010) | (0.0030) | (0.0000) | (0.0000) | (0.0000) |
| R-square             | 0.7209   | 0.7307   | 0.7424   | 0.7076   | 0.7231   |
| Prob>Chi2            | 0.0000   | 0.0000   | 0.0000   | 0.0000   | 0.0000   |
| Number of Observations | 2160   | 2767     | 2779     | 2730     | 2309     |
| Number of Groups     | 296      | 371      | 271      | 357      | 320      |

**(\*) Indicates significant at 10%**
**(\*\*)Indicates significant at 5%**
**(\*\*\*)Indicates significant at 1%**

For deeper investigating, we run regressions by grouping the data in two main periods, including the pre-crisis stage for the collected data before 2008 and the post-crisis stage for the rest of the data. The findings give some clearly evidences for changing in cross-border lending behavior of G7. First, there are higher coefficients in almost regression for the post-crisis period data, which implies that the more home bias effects since the GFC. The result is also consistent with findings in several studies (Bremus & Fratzscher, 2015; De Haas & Van Horen, 2013). Second, there are some factors are insignificant correlation with cross-border claims and others factors become important drivers for cross-border claims in post-crisis period.

Regarding geographical feature, distance is a noticeable one which has positive correlation with statistical significance for all regressions in pre-crisis, but then this variable is non-statistically significant in the post-crisis period. Additionally, G7 tend to lending money to countries which are located in the same continent in the second stage. The evidence is given by positive coefficients and statistically significant at 1% for all the regressions during that time. This result may imply that G7 lend more to remoted countries before crisis to diversify the lending portfolio. After the GFC they tend to more be home bias. Moreover, this result predicts for the death of distance after the GFC.

Interestingly, exchange rate risk is an important driver for international lending during the pre-crisis period, but it is not important during the post-crisis period. Language which is a proxy for similar culture, becomes a crucial factor affecting lending behavior after the GFC.

**Table 4.2 Regression results for the first model by pre-crisis**

|  | PROXY FOR SIMILAR AND REL | | | | |
| --- | --- | --- | --- | --- | --- |
| Variable | GDP | CREDIT | M2 | MARKET | DEPOSIT |
| _cons | -63.4102*** | -70.3107*** | -74.9957*** | -31.8083*** | -40.0593*** |
|  | (0.0000) | (0.0000) | (0.0000) | (0.0060) | (0.0030) |
| Lnsize | 16.9934*** | 19.6435*** | 21.2925*** | 7.2478** | 9.4992** |
|  | (0.0010) | (0.0000) | (0.0000) | (0.0470) | (0.0240) |
| Rel | -0.0174 |  |  |  |  |
|  | (0.7740) |  |  |  |  |
| Similar | 0.3712*** |  |  |  |  |
|  | (0.0000) |  |  |  |  |
| Rel_fd_credit |  | 0.0962 |  |  |  |
|  |  | (0.1880) |  |  |  |
| Similar_fd_credit |  | 0.2421*** |  |  |  |
|  |  | (0.0000) |  |  |  |
| Rel_fd_m2 |  |  | 0.2178** |  |  |
|  |  |  | (0.0340) |  |  |
| Similar_fd_m2 |  |  | 0.3877*** |  |  |
|  |  |  | (0.0000) |  |  |
| Rel_fd_market |  |  |  | 0.0043 |  |
|  |  |  |  | (0.9320) |  |
| Similar_fd_market |  |  |  | 0.0127 |  |
|  |  |  |  | (0.7880) |  |
| Rel_fd_deposit |  |  |  |  | -0.2106* |
|  |  |  |  |  | (0.0810) |
| Similar_fd_deposit |  |  |  |  | -0.0256 |
|  |  |  |  |  | (0.6330) |
| Lndist | 0.4642*** | 0.2283** | 0.2390** | 0.1888* | 0.3814*** |
|  | (0.0010) | (0.0360) | (0.0250) | (0.0710) | (0.0010) |
| Border | -0.3875 | -0.0308 | -0.0844 | -0.2016 | 0.0452 |
|  | (0.3740) | (0.9170) | (0.7710) | (0.4750) | (0.8840) |
| Samecontinent | 0.6327** | 0.2870 | 0.4390* | -0.0999 | 0.4911* |
|  | (0.0260) | (0.2160) | (0.0580) | (0.6560) | (0.0500) |
| Pol | -0.1713*** | -0.1834*** | -0.1699*** | -0.1746*** | -0.2018*** |
|  | (0.0000) | (0.0000) | (0.0000) | (0.0000) | (0.0000) |

| Lntrade | 0.6818*** | 0.6749*** | 0.6225*** | 0.8841*** | 0.7817*** |
|---|---|---|---|---|---|
|  | (0.0000) | (0.0000) | (0.0000) | (0.0000) | (0.0000) |
| lnFDI | 0.1445*** | 0.1612*** | 0.1599*** | 0.1668*** | 0.1458*** |
|  | (0.0000) | (0.0000) | (0.0000) | (0.0000) | (0.0000) |
| Language | 0.1529 | 0.4091** | 0.4018** | 0.4318*** | 0.2200 |
|  | (0.4420) | (0.0130) | (0.0130) | (0.0070) | (0.2350) |
| Culture | 0.0002 | -0.0020 | -0.0001 | -0.0038 | -0.0048 |
|  | (0.9640) | (0.5550) | (0.9780) | (0.2440) | (0.2080) |
| ER | -0.2401** | -0.2124** | -0.1854** | -0.2593*** | -0.2350** |
|  | (0.0160) | (0.0210) | (0.0430) | (0.0040) | (0.0140) |
| lnVIX | 0.1319** | 0.0900* | 0.0686 | 0.1237** | 0.2548*** |
|  | (0.0370) | (0.0800) | (0.1780) | (0.0140) | (0.0000) |
| R-Square | 0.7313 | 0.7302 | 0.7384 | 0.7254 | 0.7134 |
| Prob>Chi2 | 0.0000 | 0.0000 | 0.0000 | 0.0000 | 0.0000 |
| Number of Observations | 1129 | 1521 | 1527 | 1481 | 1246 |
| Number of Groups | 249 | 325 | 325 | 316 | 269 |

**(*) Indicates significant at 10%**
**(**)Indicates significant at 5%**
**(***)Indicates significant at 1%**
**Table 4.3 Regression results for the first model by post-crisis**

|  | PROXY FOR SIMILAR AND REL |  |  |  |  |
|---|---|---|---|---|---|
| Variable | GDP | CREDIT | M2 | MARKET | DEPOSIT |
| _cons | -127.7658*** | -122.9168*** | -122.6642*** | -56.3642*** | -99.8083*** |
|  | (0.0000) | (0.0000) | (0.0000) | (0.0000) | (0.0000) |
| Lnsize | 39.1216*** | 37.4542*** | 37.8445*** | 16.4027*** | 28.8370*** |
|  | (0.0000) | (0.0000) | (0.0000) | (0.0000) | (0.0000) |
| Rel | -0.0870 |  |  |  |  |
|  | (0.1490) |  |  |  |  |
| Similar | 0.8369*** |  |  |  |  |
|  | (0.0000) |  |  |  |  |
| Rel_fd_credit |  | 0.4594*** |  |  |  |
|  |  | (0.0000) |  |  |  |
| Similar_fd_credit |  | 0.6398*** |  |  |  |
|  |  | (0.0000) |  |  |  |
| Rel_fd_m2 |  |  | 0.7565*** |  |  |
|  |  |  | (0.0000) |  |  |
| Similar_fd_m2 |  |  | 0.8551*** |  |  |
|  |  |  | (0.0000) |  |  |
| Rel_fd_market |  |  |  | 0.0186 |  |
|  |  |  |  | (0.6880) |  |
| Similar_fd_market |  |  |  | 0.1033** |  |
|                    |           |           |           |           | (0.0170)  |
| Rel_fd_deposit     |           |           |           |           | -0.2925** |
|                    |           |           |           |           | (0.0210)  |
| Similar_fd_deposit |           |           |           |           | -0.3591***|
|                    |           |           |           |           | (0.0000)  |
| Lndist             | 0.2194    | 0.1098    | 0.1375    | 0.1467    | 0.1962    |
|                    | (0.1050)  | (0.3380)  | (0.2240)  | (0.1900)  | (0.1400)  |
| Border             | -0.3308   | 0.0822    | 0.0739    | 0.1146    | 0.1625    |
|                    | (0.4830)  | (0.7950)  | (0.8140)  | (0.7090)  | (0.6460)  |
| Samecontinent      | 1.5053*** | 1.1821*** | 1.4502*** | 0.6850*** | 1.4656*** |
|                    | (0.0000)  | (0.0000)  | (0.0000)  | (0.0040)  | (0.0000)  |
| Pol                | -0.2667***| -0.2965***| -0.2773***| -0.2599***| -0.4029***|
|                    | (0.0000)  | (0.0000)  | (0.0000)  | (0.0000)  | (0.0000)  |
| Lntrade            | 0.2405*** | 0.2948*** | 0.1965*** | 0.6006*** | 0.3792*** |
|                    | (0.0010)  | (0.0000)  | (0.0010)  | (0.0000)  | (0.0000)  |
| lnFDI              | 0.1427*** | 0.1279*** | 0.1165*** | 0.1219*** | 0.1228*** |
|                    | (0.0000)  | (0.0000)  | (0.0000)  | (0.0000)  | (0.0000)  |
| Language           | 0.4706**  | 0.6343*** | 0.6251*** | 0.6563*** | 0.3424    |
|                    | (0.0260)  | (0.0000)  | (0.0000)  | (0.0000)  | (0.1020)  |
| Culture            | 0.0039    | -0.0007   | 0.0031    | -0.0039   | -0.0050   |
|                    | (0.3530)  | (0.8400)  | (0.4020)  | (0.2810)  | (0.2480)  |
| ER                 | 0.0276    | 0.0732    | 0.0252    | -0.1411   | 0.0391    |
|                    | (0.8020)  | (0.4880)  | (0.8050)  | (0.1560)  | (0.7420)  |
| lnVIX              | 0.1567**  | 0.1791*** | 0.1434**  | 0.0651    | 0.2580*** |
|                    | (0.0260)  | (0.0040)  | (0.0180)  | (0.2500)  | (0.0000)  |
| R-Square           | 0.7403    | 0.7505    | 0.7556    | 0.7367    | 0.7144    |
| Prob>Chi2          | 0.0000    | 0.0000    | 0.0000    | 0.0000    | 0.0000    |
| Number of Observations | 1031  | 1246      | 1252      | 1249      | 1063      |
| Number of Groups   | 272       | 332       | 332       | 325       | 289       |

**(*) Indicates significant at 10%**
**(**)Indicates significant at 5%**
**(***)Indicates significant at 1%**

We do regression for the second model using random effects and fixed effects. The Hausman tests suggest that fixed effects are preferable.

The table 4.4 shows that most of variables present the expected signs with statistical significance. First, the size of destination countries is still important driver in pulling cross-border claims, which is showed by the positive correlation between GDP of the borrowing countries and total claims from G7. Second, public debt has a strong relationship with claims from G7, the value is negative for debt (significantly at the level of 5% in the random effects estimation). Fourth, credit has positive coefficients and significant in both random effects and fixed effects suggesting that cross-border lending has strong relations with domestic credit in the borrowing countries.

**Table 4.4 Regression results for the second model**

|  | Fixed Effects | | Random effects | |
| --- | --- | --- | --- | --- |
|  | Coefficient | P-Value | Coefficient | P-Value |
| _cons | -53.8508*** | 0.0000 | -20.9574*** | 0.0000 |
| Lngdpj | 2.1416*** | 0.0000 | 1.0077*** | 0.0000 |
| Lnexport | 0.4193*** | 0.0010 | 0.0217 | 0.8420 |
| Lnimport | 0.1271 | 0.3720 | 0.2747** | 0.0190 |
| Debt | -0.0013 | 0.5340 | -0.0050** | 0.0150 |
| Gov | -0.0358 | 0.8370 | 0.2041 | 0.1050 |
| Credit | 0.0055*** | 0.0060 | 0.0110*** | 0.0000 |
| R-Square | 0.6922 | | 0.7717 | |
| Number of Observations | 234 | | 234 | |
| Number of Groups | 53 | | 53 | |

Note: (*), (**), (***) Indicates significant at 10%, 5%, 1% respectively.

Interestingly, Table 4.5 shows that almost coefficients, except the coefficient of LnGDP, are just statistically significant with the pre-crisis data and they are non-statistically significant with the post-crisis data. This result implies that the pull factors are not important anymore after the GFC. However, only the bigger borrowing economies will attract more claims in the whole research period.

**Table 4.5 Regression results for the second model by crisis**

|  | **Fixed Effects** | | **Random Effects** | |
| --- | --- | --- | --- | --- |
|  | **Pre-Crisis** | **Post-Crisis** | **Pre-Crisis** | **Post-Crisis** |
| _cons | -88.9530*** | -37.9766*** | -14.6108*** | -23.2844*** |
|  | (0.0000) | (0.0020) | (0.0000) | (0.0000) |
| Lngdpj | 3.1974*** | 2.0232*** | 0.4720** | 1.4143*** |
|  | (0.0000) | (0.0000) | (0.0330) | (0.0000) |
| Lnexport | 0.5577** | 0.1102 | 0.0850 | -0.0156 |
| | (0.0430) | (0.4610) | (0.6480) | (0.8900) |
| Lnimport | 0.3586* | -0.3166 | 0.6369*** | -0.1882 |
| | (0.0870) | (0.1200) | (0.0000) | (0.1780) |
| Debt | 0.0117*** | -0.0050 | -0.0026 | 0.0029 |
| | (0.0080) | (0.3410) | (0.5110) | (0.4680) |
| Gov | 0.2446 | -0.0715 | 0.3581** | 0.4335** |
| | (0.2230) | (0.8370) | (0.0220) | (0.0160) |
| Credit | 0.0065** | 0.0009 | 0.0130 | 0.0026 |
| | (0.0160) | (0.7830) | (0.0000) | (0.3300) |
| R-Square | 0.4812 | 0.8213 | 0.6702 | 0.8729 |
| Number of Observations | 118 | 116 | 118 | 116 |
| Number of Observations | 36 | 36 | 36 | 36 |

Note: (*), (**), (***) Indicates significant at 10%, 5%, 1% respectively.

## 5. Conclusion

This study has investigated the cross border banking from G7 to their destination countries through the GFC. The data collected from BIS from 2001 to 2013 show that the strong fluctuation of claims is consistent with the crisis, and the trend depends on their interested destination countries. The results imply that home bias affect the financial behavior more than the diversify principle and pull factors are less important in determination of cross-border lending from G7. The finding also shows that exchange rate risk, political risk significantly discourage the claims from lenders to borrowers. Furthermore, we notice that language is an important driver since the GFC. Finally, global financial risk increases correspondingly with the amount of cross-border claims from G7, which suggests for further studies in the strong correlation between international banking activity and global financial risk.